\documentclass{article}
\usepackage{epsfig}
\usepackage{amssymb}
\usepackage{amsmath}
\usepackage{amsfonts}
\newcommand \be{\begin{eqnarray}}
\newcommand \ee{\end{eqnarray}}
\begin{document}
\begin{center}
{\bf Real-Time Kadanoff-Baym Approach to Nuclar Response Functions } \\
\bigskip
\bigskip
H. S. K\"ohler \footnote{e-mail: kohlers@u.arizona.edu} \\
{\em Physics Department, University of Arizona, Tucson, Arizona
85721,USA} \\
N.H. Kwong \footnote{e-mail: kwong@optics.arizona.edu} \\
{\em Optical Sciences Center, University of Arizona, Tucson, Arizona
85721,USA} \\
\end{center}
\date{\today}

\begin{abstract}
Linear response functions are calculated for symmetric nuclear matter 
of normal density
by time-evolving two-time
Green's functions with conserving self-energy insertions, thereby
satisfying the energy-sum rule.  Nucleons are regarded as moving in a
mean field defined by an effective mass. A two-body effective (or
residual)  interaction,  represented by a gaussian local interaction, is
used to find the effect of correlations in a second order as well as a
ring approximation.
The response function $S(\omega,q)$ is calculated for
$0.2<q<1.2 fm^{-1}$. Comparison is made with the nucleons being
un-correlated, "RPA+HF" only.

\end{abstract}
\vskip2pc

\section{Introduction\protect\\} 
This report concerns the calculation of the density response function 
in symmetric nuclear matter when subjected to an external probe.
This problem has traditionally involved solving  a Bethe-Salpeter
equation. A main problem with that method is the construction of a
consistent interaction kernel, to satisfy the energy sum-rule. 
Such a calculation was however accomplished by Bozek for nuclear
matter\cite{boz04} and further applied in ref.\cite{boz05}
Response functions have of course since long been the focus of intense
studies for the electron gas with numerous contributions. Early works
by Lundqvist and Hedin are noticeable leading to the often cited GW
method. \cite{hed69}
More recent works include those of Faleev and Stockman focussing on 
electrons in quantum wells.\cite{fal01}

An alternative method, first applied by Kwong and Bonitz for the plasma
oscillations in an electron gas, is a real time solution of the Kadanoff
Baym equations.\cite{nhk00}  With conserving approximations of the 
two-time Green's functions the energy sum rule is automatically guaranteed. 
The details of this method was already presented in the paper by Kwong
and Bonitz.
The main purpose here is to illustrate this method 
for the calculation of response functions in symmetric nuclear matter,
emphasizing the relative ease of these computations and also the
relation of these calculations to the nuclear many-body problem in
general. 

It was shown in Baym and Kadanoff's original papers \cite{bay62}
that, if one wishes to construct the linear response
function from dressed equilibrium Green's functions,
appropriate vertex corrections to the polarization bubble
are necessary to guarantee the preservation of the local
continuity equation for the particle density and current
in the excited system,
which in turn implies the satisfaction of the energy-weighted
sum rule. This condition is satisfied if the appropriate selfenergies
are calculated in a conserving approximation.

\section{Formalism}
Symmetric nuclear matter is subjected to an external
perturbation specified below.  Nucleons are initially at time $t=0$ 
assumed to be uncorrelated but moving in a mean field defined by 
an effective mass $m^{*}$. The initial temperature is usually
assumed to be $T=0$, but in some cases shown below it is assumed to be
20 MeV. In those cases where  the effect of correlations is studied an
interaction which is local (in coordinate space)
is switched on and the system is time-evolved in real time until
fully correlated, at $t=10 fm/c$. As a consequence of the conserving
approximations the system will during this time heat up. 
The external perturbation is now
turned on and the system will exhibit density-fluctuations in time.
Details are presented below.

\subsection{The two-time KB-equations}
The Kadanoff-Baym equations in the two-time form for the specific
problem at hand was already shown in previous work, where it was applied
to the electron gas.\cite{nhk00} Few modifications are necessary for the present
nuclear problem. 

We consider three separate cases: \\
I.  Uncorrelated, mean field only, HF+RPA approximation.\\ 
II. Correlations included by second order self-energies  with a
residual interaction, eq.(\ref{eq3bb}) below. \\
III. Ring-diagrams included in the selfenergy to all orders. \\ 
The selfenergy obtained with the
second order Born approximation as well as RPA
are 'conserving approximations'.  \cite{bay62}

As made clear in ref. \cite{nhk00} there is no separate need to calculate
vertex corrections. They are generated by the time-evolution of the
Green's functions.

The Green's function is separated into a spatially homogeneous part
$G_{00}$ and a linear response part $G_{01}$.

Calculations proceed as follows:
Equilibrium Green's functions are constructed for an uncorrelated fermi
distribution of specified density and temperature. These $G_{00}$ functions
are time-evolved with the chosen mean field and correlations (I,II or III
above) until stationary. (typically $10fm/c$). 
An external field $U_{0}(t)$
is then applied  
which  generates 
particle-hole functions $G^{^{>}_{<}}_{10}$ propagated by eqs (6) in
ref\cite{nhk00}. They are for completeness repeated here

\begin{eqnarray}
\left(i\hbar \frac{\partial}{\partial t}-\epsilon_{{\bf k+q_0}} \right )
G_{10}^{^{>}_{<}}({\bf k}tt')=U_0(t)G_{00}^{^{>}_{<}}({\bf k}tt')
+\Sigma_{1m}^{HF}({\bf k}t)G_{m0}^{^{>}_{<}}({\bf k}tt')
\nonumber \\
+\Sigma_{1m}^{R}({\bf k}t\bar{t})G_{m0}^{^{>}_{<}}({\bf k}\bar{t}t')+
\Sigma_{1m}^{^{>}_{<}}({\bf k}t\bar{t})G_{m0}^{A}({\bf k}\bar{t}t')
\label{eq01a}
\end{eqnarray}
and
\begin{eqnarray}
\left(-i\hbar \frac{\partial}{\partial t'}-\epsilon_{{\bf k}} \right )
G_{10}^{^{>}_{<}}({\bf k}tt')=U_0(t')G_{11}^{^{>}_{<}}({\bf k}tt')
+G_{1m}^{^{>}_{<}}({\bf k}tt')\Sigma_{m0}^{HF}({\bf k}t')
\nonumber \\
+G_{1m}^{R}({\bf k}t\bar{t'})\Sigma_{m0}^{^{>}_{<}}({\bf k}\bar{t}t')
+G_{1m}^{^{>}_{<}}({\bf k}t\bar{t'})\Sigma_{m0}^{A}({\bf k}\bar{t}t')
\label{eq01b}
\end{eqnarray}

the selfenergies $\Sigma_{00}$ are given by

\begin{eqnarray}
\Sigma_{00}^{^{>}_{<}}({\bf k},t,t')=
i\sum_{\bf p}
G_{00}^{^{>}_{<}}({\bf k-p},t,t')
V_{s}^{^{>}_{<}}({\bf p},t,t').
\label{eqring}
\end{eqnarray}

In the second order calculautions (case II)  we have

\begin{eqnarray}
V_{s}^{^{>}_{<}}({\bf p},t,t') =
V^{2}({\bf p})\Pi_{00}^{^{>}_{<}}({\bf p},t,t')
\label{eq0}
\end{eqnarray}

where $V({\bf p})$ is the momentum-representation of the residual potential,
local in ccordinate space, eq. (\ref{eq3bb}). 

The polarisation bubble $\Pi_{00}$ is defined by
\begin{eqnarray}
\Pi_{00}^{^{>}_{<}}({\bf p},t,t')=
-i\sum_{\bf p'}
G_{00}^{^{>}_{<}}({\bf p'},t,t')
G_{00}^{^{<}_{>}}({\bf p'-p},t',t)
\label{eq3c} 
\end{eqnarray}

The selfenergies in the ({10}) channel are given by
\begin{eqnarray}
\Sigma_{10}^{^{>}_{<}}({\bf k},t,t')=
i\sum_{\bf p}
[G_{10}^{^{>}_{<}}({\bf k-p},t,t')
V_{s}^{^{>}_{<}}({\bf p},t,t')     \nonumber \\
+G_{00}^{^{>}_{<}}({\bf k-p},t,t')
V_{s(10)}^{^{>}_{<}}({\bf p},t,t')]
\label{eq4}
\end{eqnarray}

with

\begin{equation}
V_{s(10)}^{^{>}_{<}}({\bf p},t,t')= 
V^{2}({\bf p})\Pi_{10}^{^{>}_{<}}({\bf p},t,t')
\label{eq7b}
\end{equation}

and the polarisation bubble in the $(10)$-channel is given by
\begin{eqnarray}
\Pi_{10}^{^{>}_{<}}({\bf p},t,t')= 
-i\sum_{\bf p'}
[G_{10}^{^{>}_{<}}({\bf p'},t,t')
G_{00}^{^{<}_{>}}({\bf p'-p},t',t) \nonumber \\
+G_{00}^{^{>}_{<}}({\bf p'},t,t')
G_{10}^{^{<}_{>}}({\bf p'-p-q},t',t)]
\label{eq3d} 
\end{eqnarray}

In case III where not only the second order but all RPA-rings are included
in the selfenrgies one has

\begin{eqnarray}
V_{s}^{^{>}_{<}}({\bf p},t,t') =
V({\bf p})[\int_{t_{0}}^{t} dt''(\Pi_{00}^{>}({\bf p},t,t'')- 
\Pi_{00}^{<}({\bf p},t,t''))V_{s}^{^{>}_{<}}({\bf p},t'',t')-
\nonumber \\
\int_{t_{0}}^{t'} dt''\Pi^{^{>}_{<}}({\bf p},t,t'')
(V_{s}^{>}({\bf p},t'',t')-V_{s}^{<}({\bf p},t'',t'))]+ \nonumber\\
V^{2}({\bf p})\Pi_{00}^{^{>}_{<}}({\bf p},t,t')
\label{eq1}
\end{eqnarray}

and 

\begin{eqnarray}
V_{s(10)}^{^{>}_{<}}({\bf p},t,t')= 
\int_{t_{0}}^{t}d\bar{t}
\int_{t_{0}}^{t'}d\bar{t'} \nonumber   \\
\{ V_s^{R}({\bf p+q},t,\bar{t})   
\Pi_{10}^{R}({\bf p},\bar{t},\bar{t'})
V_s^{^{>}_{<}}({\bf p},\bar{t'},t') \nonumber  \\
+V_s^{R}({\bf p+q},t,\bar{t})
\Pi_{10}^{^{>}_{<}}({\bf p},\bar{t},\bar{t'}) 
V_s^{A}({\bf p},\bar{t'},t')  \nonumber \\ 
+V_s^{^{>}_{<}}({\bf p+q},t,\bar{t})
\Pi_{10}^{A}({\bf p},\bar{t},\bar{t'})
V_s^{A}({\bf p},\bar{t'},t') \}
\label{eq4a}
\end{eqnarray}

where the retarded and advanced parts are given by

\begin{eqnarray}
\Sigma_{10}^{R/A}({\bf p},t,t')=\pm \theta(\pm (t-t')[\Sigma_{10}^{>}({\bf
p},t,t')-\Sigma_{10}^{<}({\bf p},t,t')]
\label{eq5a}
\end{eqnarray}
\begin{eqnarray}
\Pi_{10}^{R/A}({\bf
p},t,t')=\pm\theta(\pm (t-t')[\Pi_{10}^{>}({\bf
p},t,t')-\Pi_{10}^{<}({\bf p},t,t')]
\label{eq5b}
\end{eqnarray}
\begin{eqnarray}
V_{s}^{R/A}({\bf p},t,t')=\delta(t-t')V({\bf p})
\pm\theta(\pm (t-t')[V_{s}^{>}({\bf
p},t,t')-V_{s}^{<}({\bf p},t,t')]
\label{eq6b}
\end{eqnarray}

The Hartree-Fock selfenergy in the ($00$)-channel is approximated by
an effective mass $m^{*}=0.7m$.  In the ($10$)-channel it is given by

\begin{equation}
\Sigma_{10}^{HF}({\bf p},t)=-iV(q)\sum_{\bf p'}
 G^{<}_{10}({\bf p'},t,t)+i\sum_{\bf p'}G^{<}_{10}({\bf
 p-p'},t,t)V(p)
\label{HF}
\end{equation}

The second term, the Fock-term, was found to be  negligent  and
subsequently
omitted in calculations. The Hatree-term on the other hand is what
drives the fluctuations in the response function.

The eqs. (\ref{eq01a}) and (\ref{eq01b}) are time-evolved and the
response-function is, after  fourier-transformations of $G^{<}_{10}({\bf
p},t,t)$  and $U_{0}(t)$ with respect to time t, calculated from
\begin{equation}
S(\omega,q)=\frac{1}{\pi n_{0}U_{0}(\omega)}\sum_{\bf p}
Im G^{<}_{10}({\bf p},\omega)
\label{S}
\end{equation}
$U_{0}(t)$ was chosen to be a Gaussian with a width of $3 fm/c$.

\subsection{Interactions}
The outcome of any microscopic many-body calculation depends on the
assumption of the interactions between the individual particles.
Some information on the NN-
interaction betewen two nucleons in free space is obtained  from
scattering. It is however incomplete being only on-shell. Some off-shell
information can be obtained indirectly from bound nuclei  e.g. the
deuteron.
It has to
be complemented by theoretical input the most important of which
is the OPEP. It does however lack information on experimentally
known short-ranged
repulsions and has to be augmented by contributions from heavier
exchange particles.\cite{mac11}  There is also the question of
contributions from not only 2- but also 3- (and higher) body forces.
This is ongoing research that leaves any present nuclear many-body
calculation open for future modifications. 

With  interactions given  one can (at least in principle)
calculate properties of the 3-nucleon system as well as those of
light nuclei "exactly" by Faddeev or no-core shell-model techniques.
More generally there is however  the question of an acceptable
many-body theory.  Many-body theories usually assume
the importance of specifis terms (diagrams) in some (sometimes
unspecified) expansion resulting in some in-medium interaction such
as Brueckner's Reaction mtarix.

The theoretical problem of
nuclear structure is consequently two-fold.  \\
I. Construction of  N-N (as well as N-N-N etc) interactions. \\
II. A many body theory allowing the calculation of nuclear
properties. \\
There has been much progress relating to both of these problems
during
the last more than 50 years of nuclear research but ongoing studies 
are still intense and of course facilitated by development of computers.
We are here not implementing the full machinery of these findings
but will use the provided knowledge to model the nuclear properties.
This is of course what is mostly done in this kind of work.
Examples are the Skyrme and Gogny forces.
We shall  model the symmtric nuclear matter under
consideration here as a system of particles (nucleons) with masses
$m$ moving in a momentum dependent (Hartree-Fock) mean field interacting
with an in-medium effective 2-body interaction. In general such an
interaction would be given in momentum-space by:
\begin{equation}
<{\bf k}|K|{\bf k'}> \equiv K({\bf p,q})
\label{eqK}
\end{equation}
where ${\bf p}=({\bf k+k'})/2$ and ${\bf q=k-k'}$.
As a result of the effective interaction being non-local (in coordinate
space) the \it diagonal \rm Hartree-Fock (or "Brueckner Hartree-Fock") 
mean field is momentum-dependent which is very important. 
We therefore approximate  $\Sigma^{HF}_{00}$  by introducing an
effective mass $m^{*}=0.7m$.

In order  to find the effect of 2-body \it correlations \rm between the nucleons
moving in this mean field we approximate the in-medium
K-matrix by a simplified 'residual' interaction,
often used in this
context\cite{dan84,boz05,hsk95}.
It is local in coordinate space.
In momentum-space it reads:
\begin{equation}
V({\bf q})=\pi^{3/2}\eta^{3}V_{0}e^{-\frac{1}{4}\eta^{2}q^{2}}
\label{eq3bb}
\end{equation}
with $\eta=0.57 {\rm fm}$ and $V_0=-453$ MeV.

Notice that this interaction only
depends on the 'transverse' momentum $q$ in momentum space but not
on the momentum $p$, responsible for the concept of an effective mass.
It would therefore not be adequte for building the \it diagonal \rm
Hartree-Fock  field $\Sigma^{HF}_{00}$  specified above.

The (off-diagonal) $q$-dependence of the effective interaction is
however important in the context of the present investigation where the
response is in fact "driven" by the Hartree-term in eq. (\ref{HF}).
It  is unfortunately less accessible wheras the diagonal is more closely
related to the scattering $T$-matrix.
We shall here use  the interaction (\ref{eq3bb}) in the Hartree term of
eq.  (\ref{HF})   and for evaluating the correlation terms.

There is no inconsistency in the two choices, the one for the mean field
and the one for correlations. They are just two aspects of the sam
effective interaction.

Other possible choices would have been any of the family of Skyrme- 
or Gogny-forces  which have been used in other works related 
to the nuclear response function. They allow for the spin-isospin
dependence of the response, lacking in our choice.
Specific features of the interaction such as tensor\cite{ols04}
and pairing\cite{sed10} have been investigated. 

The main purpose of the present work is for illustration of the
2-time method. Our simple choices for the interaction fulfills this
purpose. 

\subsection{Equilibrium Temperature}

In a KB-calculation, with  the selfenergy  defined by a conserving
approximation the total energy is
conserved. Assuming, as in the present work, that the system is initially
un-correlated the  potential energy will decrease the and kinetic
energy will increase
with the same amount until the system is fully correlated and
internal equilibrium is reached. The result is an increase
in temperature from the initially set value. This situation can be
moderated by an imaginary time-stepping method.
The final, equilibrated,temperature $T_f$ will then be a function of the
initial imaginary time $\tau$ with $T_f=0$ in the limit where
$\tau \rightarrow \infty$.

For finite values of $\tau$ this final equilibrium
temperature $T_f = 1 / \beta$ and chemical potential $\mu$ is obtained
from  the following relation between $\Sigma^{<}$
and $\Sigma^{>}$ (valid at equilibrium) \cite{kad62}:
\begin{eqnarray}
\Sigma^{>}({\bf
p},\omega)=-e^{\beta(\omega-\mu)}\Sigma^{<}({\bf
p},\omega).
\label{equil}
\end{eqnarray}
Note that, as a consequence of  eq. (\ref{equil}), the ratio of the
selfenergies is independent
of momentum ${\bf p}$, which serves as a check on numerical
accuracy and that equilibrium has been reached.

With $\beta$ and $\mu$ given the equilibrium
uncorrelated distribution function is given by
\begin{eqnarray}
f(p)=1/(1+e^{\beta(\omega(p)-\mu)})
\label{ferm}
\end{eqnarray}
with
\begin{eqnarray}
\omega(p)=p^{2}/2m+Re\Sigma^{+}({\bf p},\omega)+\Sigma^{HF}(\bf p)
\label{omega}
\end{eqnarray}

The real part of $\Sigma^{+}$ is obtained from the imaginary part by
the dispersion relation, after using the relation

\begin{eqnarray}
2Im\Sigma({\bf p},\omega)=i(\Sigma^{<}({\bf p},\omega)-
\Sigma^{>}({\bf p},\omega)
\label{imsig}
\end{eqnarray}
While the uncorrelated distribution
is given by  eq. (\ref{ferm}) the correlated is different, and given by
\begin{equation}
\rho({\bf p},t)=-iG^{<}({\bf p},t,t).
\label{corrdis}
\end{equation}

 The diffence between these two distributions is important.
 If for example the initial temperature is zero in which case $f(p)$ has
 a sharp cut-off at the fermi-surface, the correlated distribution
 $\rho({\bf p})$ is 'smoothed' at the surface. The available phase-space
 for excitations differs. 

\begin{figure}
 \centerline{
 \psfig{figure=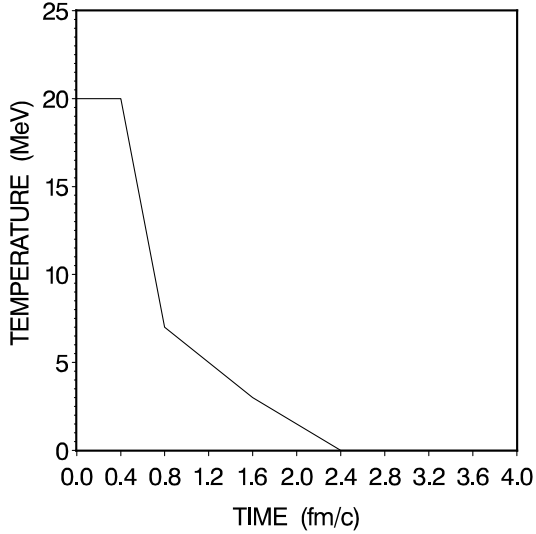,width=8cm,angle=0}
 }
 \vspace{.2in}
 \caption{The temperature as a function of imaginary time $\tau$. See
 text for details.
 }
 \label{resp1}
 \end{figure}

 Fig. \ref{resp1} shows the temperature as a function of imaginary
 time $\tau$, as calculated with the equations above.

\section{Numerical results}
\begin{figure}
\centerline{
\psfig{figure=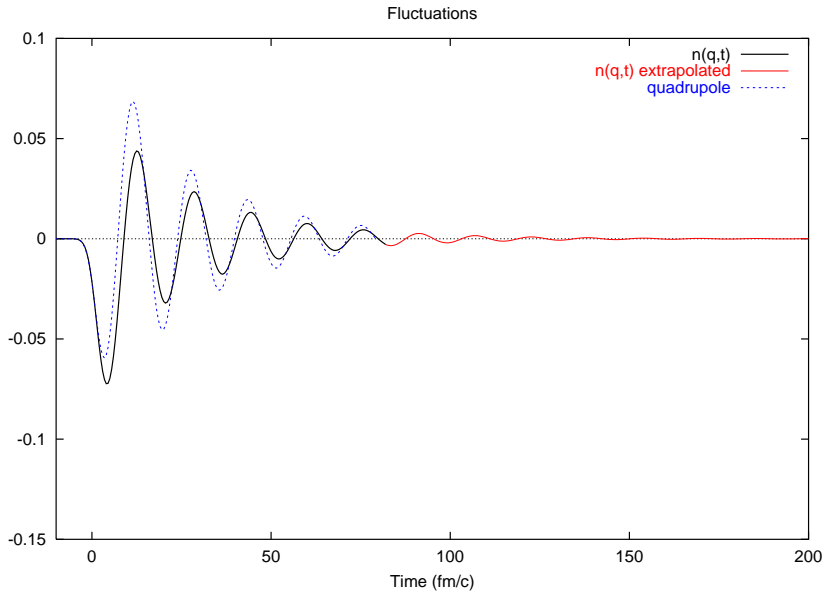,width=8cm,angle=-90}
}
\vspace{.2in}
\caption{
Solid curve shows the  fluctuating density as a function of time. It
is  in this particular case extrapolated for times $t> 82 fm/c$ 
(red online). The broken curve (blue online)  shows the associated 
quadrupole-moment of the deformation in momentum-space.
}
\label{time}
\end{figure}

Calculations are made for symmetric nuclear matter at normal nuclear matter
density, $\rho=0.16 fm^{-3}$ and some selected temperatures $T$. The
external momentum transfer is chosen to be $0.2 <q < 1.2 fm{-1}$.
The computing  method that we used was mainly as is described in ref.
\cite{hsk99} with the additional requirement to also time-evolve
$G_{01}$.\cite{nhk00} The  fluctuating density $$\delta
n(q,t)=-\sum_p G^{<}_{10}(p,t,t)$$ is recorded  until fully damped.
If the damping time is too large (number of time-steps $> \sim 150$)
then $n(q,t)$ is extrapolated using the amplitudes and frequencies
of the oscillations for lesser times. 
A typical result of the evolution in real time and an  extrapolation
is shown in Fig \ref{time}, including also the quadrupole
deformation in momentum-space. 
This time-function is then fourier-transformed to $\omega$-space.

Results for the three different 
situations referred to in the Introduction are shown below.

\subsection{Mean field +RPA; case I}
\begin{figure}
\centerline{
\psfig{figure=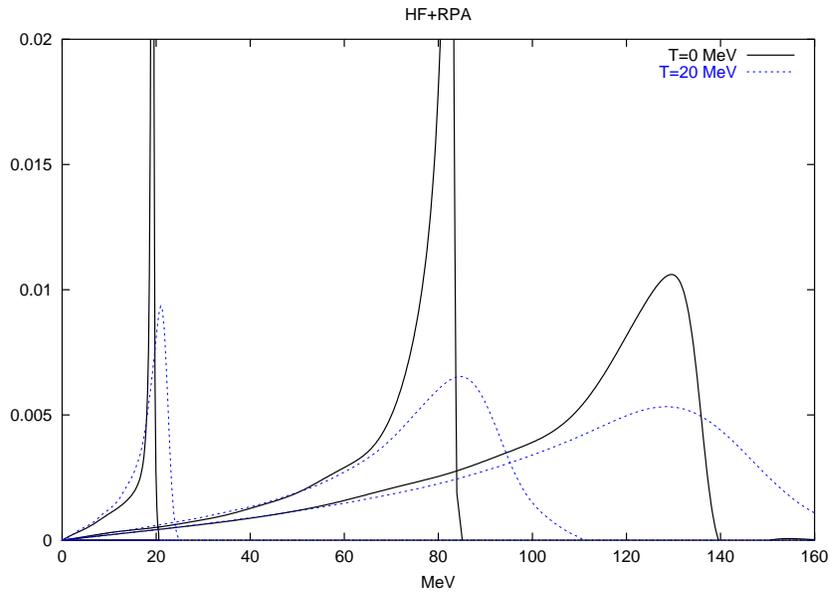,width=8cm,angle=-90}
}
\vspace{.2in}
\caption{
Response functions $S(\omega,q)$ in the HF+RPA approximation.
Solid curves show results at zero temperature for external momentum
transfers $q=0.2, 0.8$ and $1.2
fm^{-1}$ respectively. Broken curves (blue online) show results 
when temperature is raised to 20 MeV.
}
\label{resp2}
\end{figure}

The mean field is here included in an effective mass approximation with
$m^{*}=0.7$, all correlations are neglected, i.e. all selfenergies
(except the HF) are identically zero. 
Fig \ref{resp2}  show results at zero and 20 MeV temperature for
three different values of $q$.  
A sharp resonance is found for the smallest value of $q=0.2
fm^{-1}$ at temperature $T=0$. Notable is the sharp cut-off 
on the high $\omega$-side. (compare ref. \cite{mar05}).
The width increases as $q$  approaches the fermimomentum but the
sharp cut-off persists. With the
temperature increased to T=20 MeV this sharp cut-off is
eliminated and the widths at all values of $q$ is also increased.

\subsection {$2^{nd}$ order correlations; case II}
The Green's functions are in this section dressed by second order
insertions as outlined above with the interaction given by eq.
(\ref{eq3bb}).
\begin{figure}
\centerline{
\psfig{figure=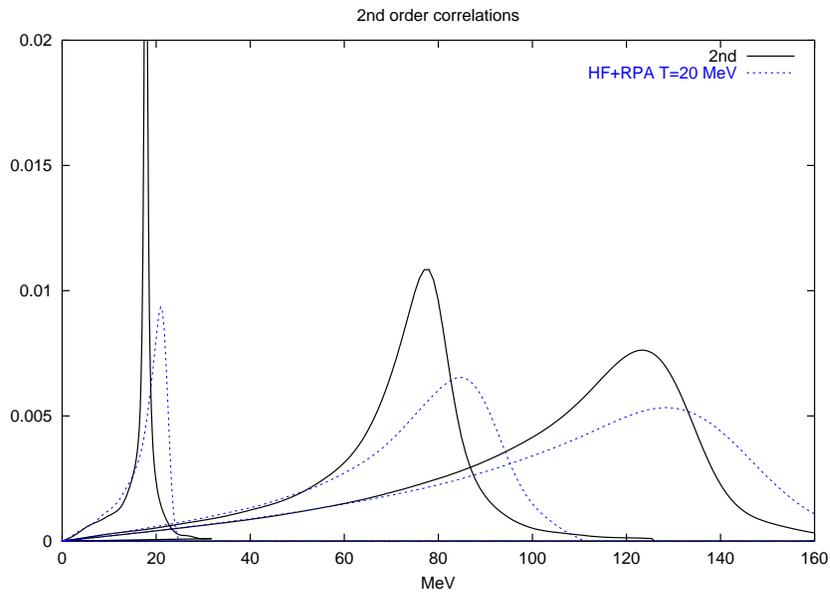,width=8cm,angle=-90}
}
\vspace{.2in}
\caption{
Solid curves show response functions $S(\omega,q)$ with $2^{nd}$ 
order correlations included.  External momentum transfers are $q=0.2, 
0.8$ and $1.2 fm^{-1}$ respectively. They are compared
with HF+RPA results at 
temperatures  T=20 MeV: broken curves (blue online). See text for further
explanations.
}
\label{resp3}
\end{figure}
The initial temperature, at time $t=0$ is here equal to zero, the
momentum-distribution is that of a zero-temperature
fermi-distribution. After the system is fully correlated (for times
$t>t_{c}$), the
temperature will as described above, have increased. Referring to
Fig.\ref{resp1} this temperature is estimated to be 20 MeV.
In the linear response limit that is used here, separating the Green's
function into $G_{00}$ and $G_{01}$ components, the collision term for
the $G_{00}$ evolution will be zero for $t>t_{c}$.
The external momentum perturbation is  applied at this time. With the
conserving approximation satisfied the vertex-correction is
'automatically' included.\cite{nhk00}
Results of these calculations are shown in Fig. \ref{resp3} for three
different values of the external momentum transfer $q$.
Comparing with the zero temperature results in Fig. \ref{resp2} one
finds no appreciable shift in resonance energies but a small broadening
that increases with $q$. The sharp cut-offs at the high energy tails are
replaced by  smoother tails, somewhat similar to what is seen in Fig.
\ref{resp2} at T=20 MeV. 
One effect associated with the correlations is a smoothing of
fermi-surface somewhat similar to that of a temperature increase. This
is of course a consequence of the broadening of the static spectral
functions caused by the imaginary parts of the self-energies.  

\subsection{Full Ring-correlations; case III}
\begin{figure}
\centerline{
\psfig{figure=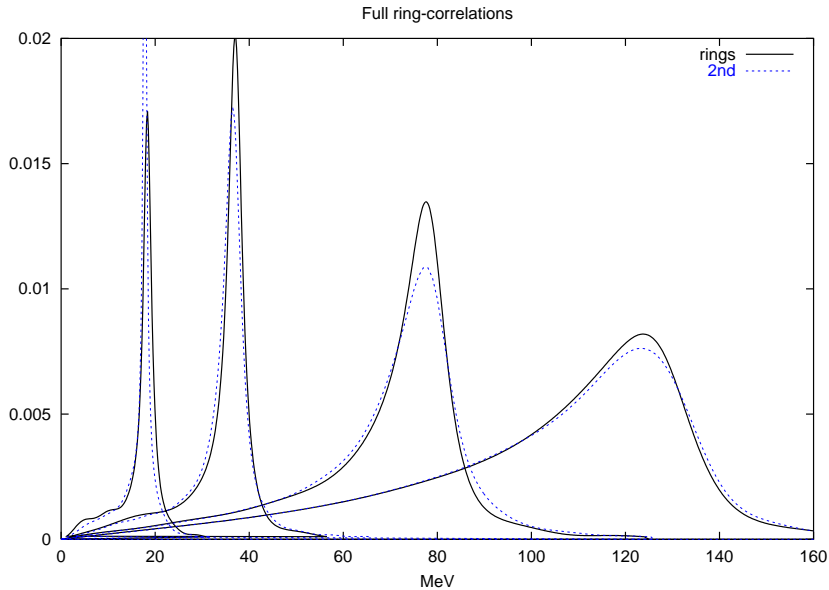,width=8cm,angle=-90}
}
\vspace{.2in}
\caption{
Solid curves show response functions $S(\omega,q)$ with full
ring correlations
included for external momentum transfers $q=0.2, 0.4,
0.8$ and $1.2 fm^{-1}$ respectively. They are compared
with $2^{nd}$ order results:  broken curves (blue online). 
See text for further explanations.
}
\label{resp4}
\end{figure}
The results above, showing the effect of including correlation to
second order in the chosen interaction is here extended with the
inclusion of polarisation bubbles to all orders. This seems a logical
extension of the second order which after all is just the lowest order
of the full ring expansion. This is of course well known to be of
extreme importance in the theory of the electron gas, providing
screening. Fig. \ref{resp4}
shows our result. The difference from the $2^{nd}$ order, that is shown
by the broken line, is seen to be small. Only a slight sharpening of
the resonances is observed.

\section{Summary and Conclusions}
The main purpose of this presentation is to illustrate the usefulness
of the two-time Green's function (Kadanoff-Baym) method for the study
of nuclear response. It was previously applied to the electron gas to
see the effect of correlations.\cite{nhk00}  The time-evolution of
Green's function in two-time space with conserving approximations for
the self-energies guarantees the preservation of sum-rules. An
alternative method, the solution of the Bethe-Salpeter (BS) equation (in
$\omega$-space) has been carried through by Bozek et
al.\cite{boz04,boz05}.
It is found that zero temperature and/or small $q$ 
calculations, where the spectral width is
small, present numerical problems when using this BS-formalism. 

This is not a problem when using the present method  
with the two-time Green's functions.

The rise in temperature associated with the onset of correlations when
using the two-time method is sometimes cited as a drawback. It is
however not difficult to remedy by allowing the correlations to
proceed in the imaginary time-domain. It requires of course to have
access to the proper computer-codes. It is our aim to utilise these
existing codes in future work.

Comparing the results obtained so far in our on-going investigation
one does not find any major difference in the three different
approximations, collisionless, $2^{nd}$-order or full rings. 
In this regard our result agrees with those of ref. \cite{boz05}.

The method of two-time Green's function time-evolution is well
documented, but there are certainly improvements in its application to 
the problem of nuclear response that are desired. Most of these are
common with any nuclear many body problem such as NN-forces and
in-medium effects.
Although our understanding of nuclear forces and of the nuclear many-body
problem is under constant development it is till incomplete.  There is 
however important knowledge
 that can and should  be incorporated in future nuclear response
calculations. Spin-isospin, tensor and pairing are all aspects of the
interaction that
have already been shown to be of interest\cite{boz05,ols04,sed10}
and still remain to be included in our calculations.

The results shown are for symmetric nuclear matter. Response calculations
for neutron matter is of particular interest related to astrophysical
problems. 

The local interaction that has been used to find the effect of
correlations may be adequate but a more "realistic"
non-local,state-dependent,
interaction should be used. Separable interactions have been
developped from inverse scattering that can easily be incorporated in
our calculations. The long-range ($V_{low-k}$) version of these would
be adequate. It has already ben shown that as far as the spectral
width concerns it is indeed the long-ranged part that plays the
important role.\cite{hsk66}
The self-energy insertions have been separated into a Hartree-Fock (mean
field)
that is real and a 'correlation' part which is complex. This can be done
consistently in the case of weak interactions. With medium-dependent
effective interactions such a separation is not well defined and can
lead to double-counting. In the
present work the mean field is assumed to be included by an effective
mass that could for example be the result of a Brueckner calculation
involving a summation of ladder diagrams and more. The real part of the
second order self-energy insertion used to find the effect of
correlations would therefore already be included in a Brueckner  mean field.
So one may claim that a double-counting has been done.
The main objective of the second order insertion is however to find the
effect of the broadening of the spectral-function and this is a
consequence
of the insertion having an imaginary part and, as mentioned, this
imaginary part stems mainly from the
long-ranged part of the interaction. As a result we claim that the
important physics is (reasonably well) incorporated in our 
calculations. Future work should however address this point in a fully
consistent way.
More important is however a fully consistent calculation of the
$q-dependent$ Hartree field that is the driving mechanism of the response.
It was derived from the same local
interaction as for the second order selfenergy. An better alternative would
be for  example to use the Brueckner $K$-matrix as outlined above in the
Introduction.

\section{Acknowledgements}

One of us (HSK) wishes to thank The University of Arizona and in
particular the Department of Physics for providing office space and
access to computer facilities.
	This work has not been supported by any external agency.


\begin{thebibliography}{10}
\bibitem{boz04}  P. Bozek,
                 Phys.  Letters {\bf B579} 309 (2004).
\bibitem{boz05}  P. Bozek,J. Margueron and H. M\"uther, 
                 Ann. \ Phys.(N.Y.) {\bf 318} 245 (2005).
\bibitem{hed69}  L. Hedin and S. Lundqvist,
                 Solid State Physics (Acadenmic, New York, 1969) Vol
		 23.
\bibitem{fal01}  Sergey V. Faleev and Mark.I. Stockman
                 Phys. \ Rev.B {\bf 63} 193302 (2001),
                 Phys. \ Rev.B {\bf 66} 085318 (2002).
\bibitem{nhk00}  N. H. Kwong and M. Bonitz, 
                 Phys. \ Rev.\ Letters {\bf 84} 1768 (2000).
\bibitem{bay62} Gordon Baym and Leo P. Kadanoff,
                 Phys. \ Rev. {\bf 124} 287 (1961) ;  Gordon Baym,
                 Phys. \ Rev. {\bf 127} 1391 (1962).
\bibitem{kad62} L.P. Kadanoff and G.Baym, Quantum Statistical Mechanics.
               (Benjamin,New York, 1962).  
\bibitem{mac11} R. Machleidt and D.R. Entem, 
                Physics Reports {\bf 503} 1 (2011).
\bibitem{dan84}  P. Danielewicz, Ann. \ Phys.(N.Y.) {\bf 152} 305 (1984).
\bibitem{hsk95}  H.S. K\"ohler, Phys. \ Rev.C {\bf 51} 3232 (1995).
\bibitem{ols04} E. Olsson, P. Haensel and C.J. Pethick,
                 nucl-th/0403066,
		 G.I. Lykasov, E. Olsson and C.J. Pethick,
                 nucl-th/0502026
\bibitem{sed10}  Armen Sedrakian and Jochen Keller,
                 nucl-th/1001.0395
		 Jochen Keller and Armen Sedrakian
                 nucl-th/1205.6902
\bibitem{hsk99} H.S. K\"ohler, N.H. Kwong and Hashim A. Yousif,
                Comp.Phys.Comm. {\bf 123}  123 (1999).
\bibitem{mar05} J. Margueron, Ngguyen Van Giai and J. Navarro,
                 nucl-th/0507053
\bibitem{hsk66} H.S. K\"ohler 
                Nucl. Phys. {\bf 88} 529(1966) 
\end{thebibliography}
\end{document}